\documentclass[pra,twocolumn,aps]{revtex4}
\usepackage{graphicx}  
\usepackage{dcolumn}   
\usepackage{bm}        
\usepackage{amssymb}   
\usepackage{amsmath}
\usepackage{setspace} 

\begin{document}

\title{Correlation spectroscopy in cold atoms: light sidebands resonances
in  electromagnetically induced transparency condition}

\author{H. M. Florez$^{1}$,  A. Kumar$^{1}$, K. Theophilo$^{1}$, P. Nussenzveig$^{1}$ and M. Martinelli$^{1}$}

\affiliation{$^{1}$Instituto de F\'{\i}sica, Universidade de S\~ao Paulo, 05315-970 S\~ao Paulo, SP-Brazil}

\begin{abstract} 
The correlation spectroscopy has been successfully employed in the measurement of the intrinsic linewidth of electromagnetically induced transparency (EIT) in time and frequency domain. 
We study the role of the sidebands of the intense fields in the measured spectra, analyzing  the information that can be recovered working with different analysis frequencies. 
In this case, the non-zero one-photon detuning appears as a necessary condition for spectrally resolving the sideband resonances in the correlation coefficient. 
Our experimental findings are supported by  the perturbative model defined in the frequency domain. 
\end{abstract}

\maketitle



One of the widely used mechanisms that provide coherent mapping between light and matter is  electromagnetically induced transparency (EIT) ~\cite{Harris,HarrisExp,Marangos05}. Such light-matter interface finds applications in quantum memories and quantum repeaters ~\cite{Duan2001,Lukin03,Polzik10}. It can also be employed for quantum non-demolition measurements \cite{Grangier97} and as a source of intense quantum correlated light beams \cite{Dantan06, Garrido03}. 

Recently, it was shown that the correlation spectroscopy can be used as a technique for measuring the intrinsic linewidth of the EIT resonance \cite{Xiao09,Felinto13,HMFlorez13}, which is directly related to the decoherence lifetime of the involved ground states. This linewidth is narrower than the broadened EIT linewidth usually measured by the standard transmission spectroscopy.
Therefore, it was shown to be a useful spectroscopic tool to estimate the decoherence limitations in quantum memories protocols based on EIT. These intensity correlation techniques could also be of interest in precision measurements in atomic clocks relying on coherent population trapping  \cite{Vanier05}.

Different research groups have been investigating how intensity correlation of two fields are affected by a medium, with special interest in the EIT process. Noise spectroscopy have shown to be a powerful method for this investigation \cite{Yabusaki91}. The analysis of the intensity correlation can be done either  in time  \cite{Scully05,  Scully10, Felinto13} or in frequency domains \cite{Garrido03,Martinelli04,Cruz07,Scully08, Xiao09, HMFlorez13}, using respectively the $g^{(2)}(\tau)$ function or its Fourier transform $C(\omega)$.

Felinto \textit{et al}.~\cite{Felinto13} proposed a heuristic model for 
$g^{(2)}(0)$, and applied it to cold atomic systems where Doppler broadening is negligible. It was demonstrated that correlation noise spectroscopy is power broadening free in this case, and that the linewidth of the $g^{(2)}(0)$ function is determined by the ground state decoherence lifetime.

In the frequency domain, the power broadening free features of the intrinsic EIT linewidth have been shown in a study of intensity correlation spectra $C(\omega)$ obtained with cold alkali atoms (Rubidium and Cesium)~\cite{HMFlorez13}, with the advantage that with proper selection of analysis frequency the contribution from electronic noise to the measurement is either negligible or can be easily subtracted. 
Moreover, with the recent developments of quantum optics in frequency domain (specially multimode quantum optics \cite{Fabre12}, and the observation of multi-mode coherent effects \cite{Campbell2009})
 stimulates the investigation of the atomic behavior in similar situations.
 
Careful study of correlations in frequency domain shows the contributions from the sidebands of the main field 
 \cite{HMFlorez15}. While they don't affect the narrow structure related to the intrinsic linewidth, the whole profile of the correlation inside the typical  EIT linewidth is broadened by the sidebands resonances, a major difference between analysis in time \cite{Felinto13} and frequency \cite{HMFlorez13} domains.

In this work, we explore the role of the sideband resonances and the effect of the detuning in the correlation profile. 
 We show that the interplay between one-photon detuning 
 and analysis frequency $\omega$ opens the possibility of detecting atomic response to sidebands  in correlation spectra. 
We also show that in such condition the correlation linewidth measured in the frequency domain is exactly the same as the one in the time domain for the $g^{2}(0)$ described in \cite{Felinto13}.
Moreover, it is shown experimentally that when one of the beams is kept in resonance with the atomic transition,  independent correlation information of the sidebands and the carriers is unavailable~\cite{HMFlorez15}, therefore demonstrating the richer structure contained in $C(\omega)$ in comparison with $g^{2}(0)$.

In addition, the perturbative method we use provides a physical insight on the mapping of the atomic response with the correlation functions $g^{(2)}(0)$ and $C(\omega)$. Unlike the resonant case, where the dispersive response seems responsible for the main contribution in the intensity correlation, in the non-resonant case there is a balance between the absorptive and dispersive response. Such a balance makes the heuristic model for $g^{(2)}(0)$ a valid approximation of $C(\omega)$ only inside the typical EIT linewidth, failing at higher two-photon detunings .



Our presentation is organized as follows. In section \ref{Theory}, we briefly describe the technique of correlation spectroscopy and the level scheme used in our study. We also describe the theoretical model used to define the correlation coefficient in frequency domain. In section \ref{Experiment}, we discuss the details of our experiment, starting from a cold cloud of $^{87}$Rb, and present the control of the involved parameters. In section \ref{Results}, we show the results of correlation spectroscopy with different values of one photon detuning, and compare it to the theory.
We also show how the presence of the sidebands is revealed in the correlation spectrum obtained at different one photon detunings.  In section \ref{Discussion}, we discuss our results and in section  \ref{Conclusions} we summarize our findings of correlation spectroscopy done at different values of one photon detuning. 

\section{Theory: Correlation spectroscopy}\label{Theory}

Inspired in the Hanbury-Brown and Twiss's experiments, the intensity correlation between two light fields with intensities $I_1(t)$ and $I_2(t+\tau)$ can be quantified by the $g^{(2)}(\tau)$ function \cite{Scully05}. 
Our interest is to analyze such intensity correlation between two light fields induced by cold atomic media in EIT condition. Yabuzaki \textit{et al}. \cite{Yabusaki91} showed  that the atomic medium converts excess phase noise of the input light sources into intensity noise at the output field. Therefore, to model the intensity correlation induced by the atoms in our bipartite system, the transformation of phase noise to amplitude noise has to be considered.

\begin{figure}[htb!]
\centering
\includegraphics[width=86mm]{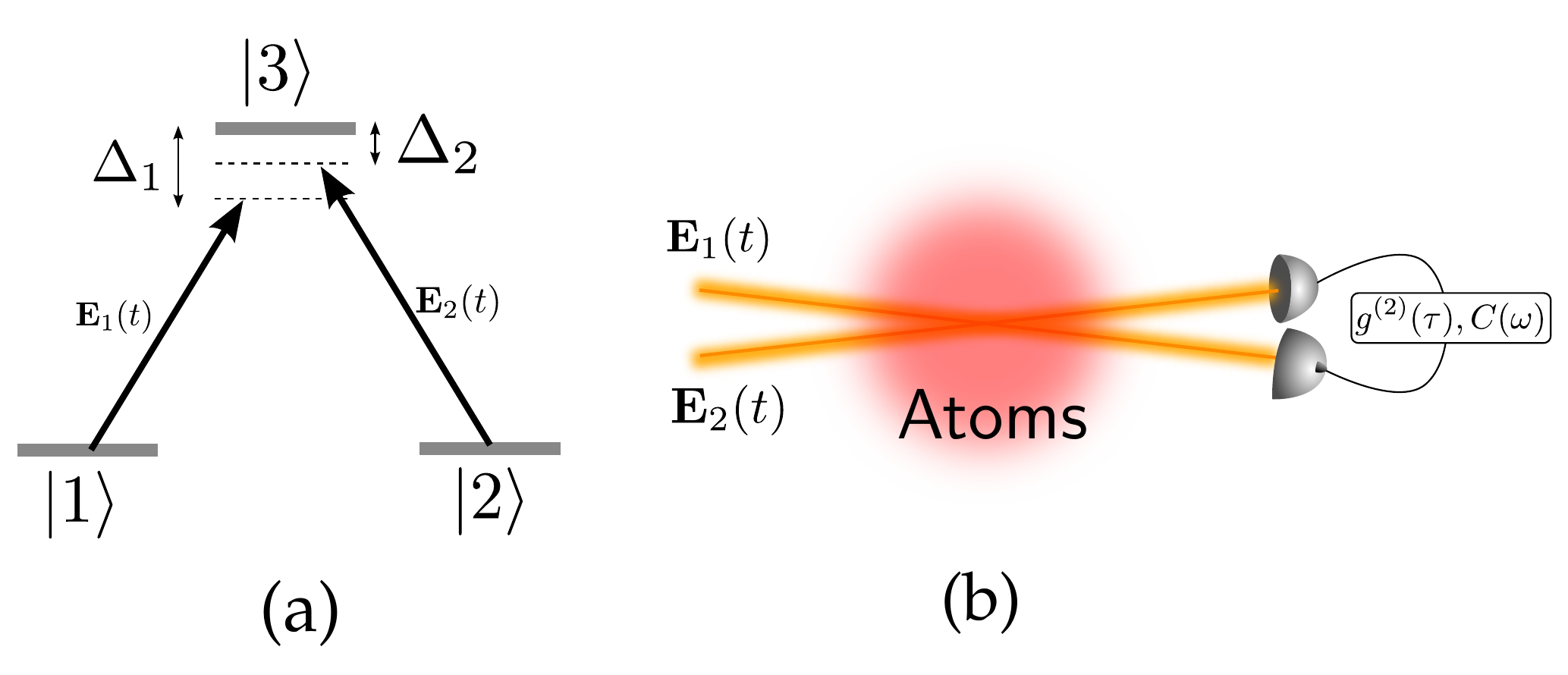}
\vspace{-.6cm}
\caption{(Color online) (a) Levels scheme in $\Lambda$-type configuration for EIT. (b) Basic setup for measuring the intensity correletion spectra in time ($g^{(2)}(\tau)$) or frequency domain ($C(\omega)$).}
\label{fig:Levels}
\end{figure}

Our scheme for correlation spectroscopy in a $\Lambda$- EIT condition is shown in Fig.~\ref{fig:Levels}(a).  Two fields,  $\mathbf{E}_{1}(t)$ and  $\mathbf{E}_{2}(t)$ are coupled to two different transitions of a three-level atom, with different one-photon detunings $\Delta_1$ and $\Delta_2$ respectively. The intensity correlation between the two fields is measured after interacting with the atomic ensemble as presented in Fig.\ref{fig:Levels}.(b). The previously reported correlation spectroscopy ~\cite{Xiao09,Felinto13,HMFlorez13, Xiao14} has been done by setting one of the fields in resonance (e.g. $\Delta_1=0$), while scanning the detuning of the second field around the resonance. 
However,  in the present work, we set different values of detuning $\Delta_1$ i.e.  $\Delta_1\neq 0$ while $\Delta_2$ is scanned around $\Delta_1$.

In what follows, we present a brief description of the semiclassical approach for the intensity correlation spectrum  based on the conversion of phase noise to amplitude noise. We consider two electromagnetic fields described by 
\begin{equation}
\mathbf{E}_{i}(t)={\cal E}_{i} \exp \left[i(\omega_{i} t+ \phi_{i}(t))\right] \mathbf{e}_{i},
\label{Efield}
\end{equation}
with a stochastic phase fluctuation $ \phi_{i}$ that models  the excess of noise in diode lasers. In the expression for the fields, $i=1, 2$ denotes the two beams, ${\cal E}_{i}$ and $\omega_{i}$ are respectively the amplitude and the frequency of each beam. 

Photodetectors are not sensitive to the phase noise of light beams. However, the interaction of light fields through an atomic medium makes it detectable by mapping the phase noise into amplitude noise (PN-AN). Under the thin sample limit~\cite{Zoller94}, the field after propagation through atomic media is given by $\mathbf{E}_{i}^{out}(t) = \mathbf{E}_i(t) + i \kappa\mathbf{P}_i(t)$, where polarization $\mathbf{P}_i(t)$ represents the atomic response induced by the incident fields $\mathbf{E}_i(t)$ and $\kappa_i$ is a real constant that depends  on the atomic density. The induced polarization term is responsible for transforming the phase noise $\phi_{i}$ (contained in $\mathbf{E}_i(t)$) into amplitude noise detected in $\mathbf{E}_{i}^{out}(t)$. In a $\Lambda$-EIT configuration (Fig.~\ref{fig:Levels}a), the  polarization induced in atomic medium by two incident fields can be given as $\mathbf{P}_i= \mathbf{d}_{i3} \rho_{i3}$, where $\mathbf{d}_{i3}$ and $\rho_{i3}$ represent the electric dipole moment and the atomic coherence associated to the fields $i=1,2$. The output intensity of the beam is given by $I_{i}^{out}(t)=|\mathbf{E}_{i}^{out}(t)|^2$. 

In order to describe the correlation between two intense light beams,  we average correlations of the fluctuations in both beams, $\delta I_i(t)=I_i^{out} (t)- \langle I_i^{out}\rangle$ with $i=\{1,2\}$
 for different instants of time separated by $\tau$. There are two possible approaches to calculate the intensity correlation: in  time domain approach adopted in \cite{Scully08,Scully10,Xiao09,Felinto13}, the intensity correlation between two light fields can be easily described in terms of the atomic variables as~\cite{Felinto13}, 
\begin{eqnarray}
g^{(2)}(0)&=&\frac{\langle\delta I_1(t)\delta I_2(t)\rangle}{\sqrt{\langle\delta ^2 I_1(t)\rangle\langle\delta ^2I_2(t)\rangle}}\nonumber\\
 &=&\frac{\text{Re}\ p_1 \ \text{Re}\ p_2 +\text{Im}\ p_1 \ \text{Im}\ p_2 }{\sqrt{(\text{Re}^2 p_1+\text{Im}^2 p_1)(\text{Re}^2 p_2+\text{Im}^2 p_2)}}, \label{g20}
\end{eqnarray}
where polarization terms are related to atomic coherences as $p_i= \rho_{i3}^{ss}$ and the notations $\text{Im}^2p_i=(\text{Im }p_i)^2$ and  $\text{Re}^2p_i=(\text{Re } p_i)^2$. The $ss$ index stands for steady-state. The atomic response induces correlation ($g^{(2)}(0)>0$) or anticorrelation ($g^{(2)}(0)<0$) between the light fields, depending on two photon detuning $\delta=\Delta_1-\Delta_2$ ~\cite{Scully08,Scully10,Xiao09}. 
At low power regime, absorption dominates and the medium induces correlation between the light fields. However, as the power of the beams are increased, the dispersion term $Re\langle p_1\rangle \ Re\langle p_2\rangle<0$ overcomes the contribution from the absorption term $Im\langle p_1\rangle \ Im\langle p_2\rangle>0$, leading to anti-correlated light fields.

Other possible approach is the analysis of intensity fluctuations $\delta I_i(t+\tau)$  in their different spectral components by a Fourier transform. Therefore, the correlation in the frequency domain description  is defined as
 \begin{equation}
C(\omega) = \frac{S_{12} (\omega)}{\sqrt{S_{11}(\omega)S_{22}(\omega)} },
\label{eq:corfreq}
\end{equation}
where $S_{ij} (\omega)$ represents the symmetrical intensity correlation spectrum for  $i$ and $j$ fields at a given analysis frequency $\omega$ such that
\begin{align}
 S_{ij}(\omega)  =\frac{1}{4\pi}\int_{-\infty}^\infty d\tau e^{-i\omega\tau}& [\langle \delta I_i(t)  \delta I_j(t+\tau) \rangle\nonumber \\ &+\langle \delta I_j(t)  \delta I_i(t+\tau) \rangle].
 \label{SpectrInt}
 \end{align}
Similar to $g^{(2)}(0)$, the function $C(\omega)$ is also normalized such that for correlated fields  $1\geq C(\omega)>0$, and for anticorrelated fields $1\leq C(\omega)<0$. 

A perturbative approach is proposed in ref. \cite{HMFlorez15} to describe the correlation coefficient of eq.(\ref{eq:corfreq}) in terms of the absorption and dispersion of the fields, similar to the case of  $g^{(2)}(0)$ function in eq.(\ref{g20}). Taking the laser linewidth $\gamma$ as perturbative parameter $\epsilon$ in the expansion of the spectral components of the fluctuations, we can obtain a direct expression for the correlation $C(\omega)$, at the lowest order. The resulting expression, 
\begin{equation}
C(\omega)=\frac{\nu_{Im}(\omega)\tilde{\Pi}_{Im}+\nu_{Re}(\omega)\tilde{\Pi}_{Re}+\tilde{\Pi}_{cross}(\omega)+C_1 }{\sqrt{S_{11}(\omega)S_{22}(\omega)}}, \label{CPert}
\end{equation}
where the absorptive, dispersive and cross terms are respectively defined by the polarization terms associated with the atomic coherences
\begin{subequations}
\begin{align}
\tilde{\Pi}_{Im}&=2\epsilon^2\text{Im}\ p_1\ \text{Im}\ p_2+O(\epsilon^4)\\
\tilde{\Pi}_{Re}&= 2\epsilon^2\text{Re}\ p_1\ \text{Re}\ p_2+O(\epsilon^4)\\
\thickspace\tilde{\Pi}_{cross}(\omega)&=\nu_{RI}(\omega)[-2\epsilon^2\text{Im}\ p_1\ \text{Re}\ p_2+O(\epsilon^4)]\\
&+\nonumber \nu_{IR}(\omega)[-2\epsilon^2\text{Im}\ p_2\ \text{Re}\ p_1+O(\epsilon^4)]\\
&=\nu_{RI}(\omega)\tilde{\Pi}_{RI}(\omega)+\nu_{IR}(\omega)\tilde{\Pi}_{IR}(\omega)
\end{align}\label{Pi_transform1}
\end{subequations}
Two main diferences will show up when we compare $C(\omega)$ (eq.~\ref{CPert}) with $g^{(2)}(0)$ (eq.~\ref{g20}). One is the presence of a cross term involving the product of dispersive and absorptive contributions,  which is absent in $g^{(2)}(0)$. The other is the need of weighting coefficients  $\nu_{Im}(\omega)$, $\nu_{Re}(\omega)$ and $\nu_{RI}(\omega)$($\nu_{IR}(\omega)$) distorting the main contributions for the correlation (absorption, dispersion and Stokes transitions respectively) with a dependence on the analysis frequency $\omega$. 
A description of the calculation, as well as the definition for the smaller term $C_1$, and the explicit expressions for $S_{11}(\omega)$ and $S_{22}(\omega)$ are  given in Appendix \ref{apendA}.

Thanks to this perturbative approach, we will understand the observed anticorrelation for 
$|\delta|>\Gamma$ when $\Delta_1=0$ as a consequence of dispersion overcoming absorption. Such situation is very different from the  behavior of $g^{(2)}(0)$ functions where the absorptive and dispersive properties contribute equally to the correlation \cite{Felinto13}. It can also be understood the role of the mixing of sidebands resonances  in the PN-AN conversion when the carrier frequency $|\delta|<\Gamma$.  This approach is  an interesting resource  for finding situations where the sidebands resonances are independent of the PN-AN mechanism when the carriers are near the two-photon resonance, and compare it to the experiment.


\section{Experiment}\label{Experiment}

\begin{figure}[htb!]
\centering
\includegraphics[width=86mm]{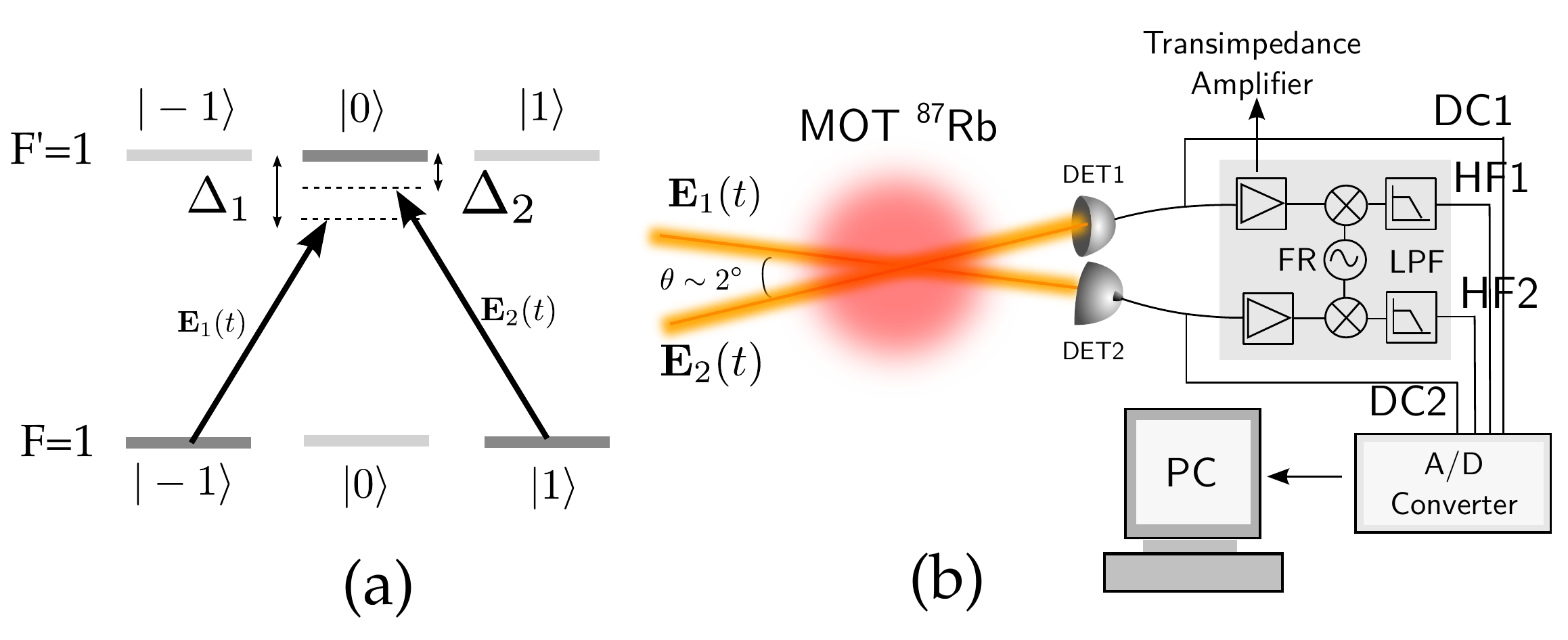}
\vspace{-.6cm}
\caption{(Color online) (a) Levels scheme in $\Lambda$-type configuration for EIT, detunings $\Delta_1$ (fixed) and $\Delta_2$ (scanned around $\Delta_1$). (b) Basic setup for EIT spectroscopy in cold atoms, where DC and HF signal are detected separately for each beam.}
\label{fig:Correlations}
\end{figure}

We used an ensemble of cold atoms of $^{87}$Rb released from a Magneto-Optical Trap (MOT). Diode lasers in $5S_{1/2}(F=2)\leftrightarrow  5P_{3/2}(F'=3)$ transition were used for cooling the atoms. Our atomic cloud has nearly $10^7$ atoms at 800$\mu$K. The optical depth of the cloud in the cooling transition is $\sim$2. The $\Lambda$-type three level system considered in the experiment is shown in Fig.~\ref{fig:Correlations}a. The correlation spectroscopy is done in  the transition $F_G=5S_{1/2}(F=1)\rightarrow F_E= 5P_{3/2}(F'=1)$ where the two fields with orthogonal polarizations $\sigma \mp$ couple the Zeeman levels $m=\pm1$ respectively, as shown in Fig.~\ref{fig:Correlations}(a). To bring the atomic population to  $5S_{1/2}(F=1)$, we used an additional optical light field coupling $5S_{1/2}(F=2)\rightarrow  5P_{3/2}(F'=2)$ transition. 
Moreover, the proposed atomic transitions obey the necessary condition $F_G \geq F_E$ for the observation of EIT in the atomic medium~\cite{Arturo99a}.

The two beams with orthogonal polarization components come from the same external cavity diode laser. Although the laser has a linewidth of $\sim$1MHz (consistent with the stochastic phase noise), the fields are phase correlated to the limit of the Standard Quantum Level (SQL). Each light beam is switched and frequency controlled by independent Acousto Optical Modulators (AOM) in double pass configuration. The frequency detuning of each beam is controlled by a computer controlled Digital-to-Analog interface acting on the AOM drivers through voltage controlled oscillators. The optical power for each beam was $P_1=112(1)\mu$W and $P_2=107(1)\mu$W, with spot sizes of $w_1=1.8(4)$mm and $w_2=1.6(4)$mm respectively (with $w$ as the spot radius at $1/e^2$ intensity level of a Gaussian beam). Therefore, the beams are below saturation, with $I_1/I_{sat}=0.18$  and $I_2/I_{sat}=0.22$. As shown in Fig.~\ref{fig:Correlations}b (not to scale), the two light beams interact with the atomic ensemble at an angle of $\sim 2^{\circ}$ to avoid any leakage from one beam to other in the detection process. 

We used \textit{p-i-n} photodetectors where the photocurrent signals are divided into a DC signal and a High Frequency (HF) signal that is connected to a trans-impedance amplifier. Both the signals are sent to an Analog-to-Digital converter and the data is acquired in a computer. The intensity noise correlation is measured from the HF signal of both detectors. The time sequence for synchronizing the MOT, the probe beams for spectroscopy and the detection scheme is the same as described in ref.~\cite{HMFlorez13}. The spectroscopy is performed in 0.5 ms while scanning the detuning of beam  2, detuning of beam 1 is kept at a fixed value, and all presented spectra are averaged over 100 scans.

\section{Results}\label{Results}

\begin{figure}[hbt!]
\centering
\includegraphics[width=180.5mm]
{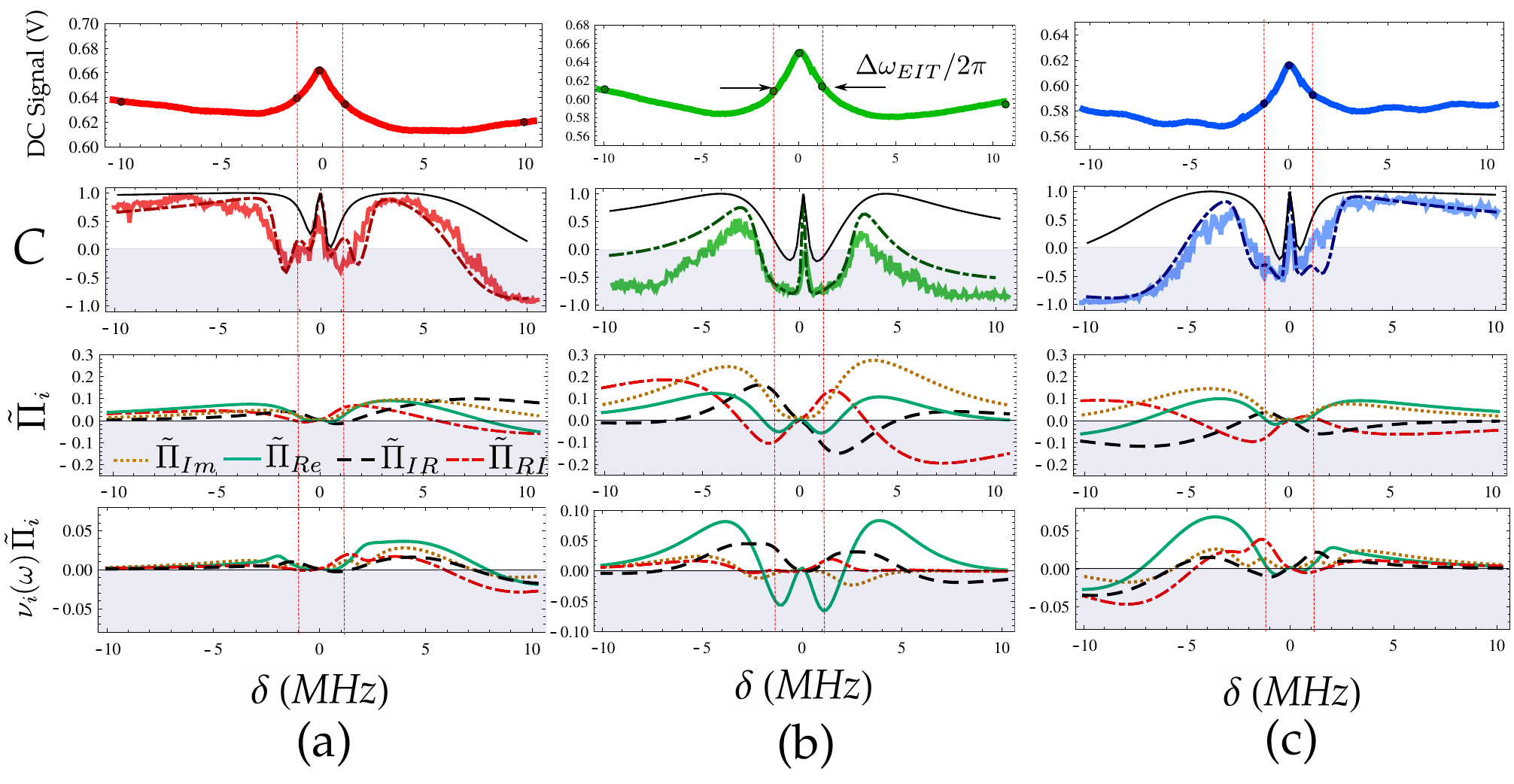}
\vspace{-0.cm}
\caption{(Color online) Spectroscopic results with three different features: DC signal (first row),  correlation (second row),  $\tilde{\Pi}$ matrix elements (third row) and $\nu(\omega)\tilde{\Pi}$ coefficients (fourth row).  Spectroscopy for (a) detuned case $\Delta_1=-6$MHz  (b) resonant case $\Delta_1=0$MHz and (c) detuned case $\Delta_1=5$MHz.  The experimental correlation spectra  in (a), (b) and (c) are shown in solid thick lines. The perturbative result for $C(\omega)$ and heuristic model for $g^{(2)}(0)$ are trace in dotted-dashed and solid thin lines respectively (second row).
According to eqs. (\ref{Pi_transform1}) the matrix elements  of  $\tilde{\Pi}$  are $\tilde{\Pi}_{Im}=2\epsilon^2 \text{Im}p_1\text{Im}p_1+\cdots $,  $\tilde{\Pi}_{Re}=2\epsilon^2 \text{Re}p_1\text{Re}p_1+\cdots $, $\tilde{\Pi}_{IR}=- 2\epsilon^2 \text{Re}p_1\text{Im}p_2+\cdots $ and $\tilde{\Pi}_{RI}=-2\epsilon^2 \text{Re}p_2\text{Im}p_1+\cdots $.
The parameters used for the atomic medium are: decoherence rate $\gamma_d/2\pi=150$kHz, analysis frequency $\omega/2\pi=2$MHz, phase noise $\bar{\gamma}/2\pi=1$MHz. The Rabi frequencies corresponding to the experimental values of intensity are $\Omega_1=0.30\ \Gamma$ and $\Omega_2=0.34\ \Gamma$. The one photon detunings used for the theoretical spectra are $\Delta_1=-6$MHz, $0.2$MHz and $5$MHz respectively from left to right. }
\label{fig:dcHFCorr_signal}
\end{figure}

In Fig.~\ref{fig:dcHFCorr_signal}, we show the spectroscopic results for three different values of one-photon detuning $\Delta_1$:  $-6$MHz (a), $0$MHz (b) and $5$MHz (c).  Each row in Fig.~\ref{fig:dcHFCorr_signal} represents different features. The top row presents the DC signal, which measures the transmission of the beams and is associated with the mean value of the photocurrent. The second row presents the correlation function $C(\omega)$ for an analysis frequency $\omega/2\pi=2$MHz;. The third and fourth row show the calculated values of the main $\tilde{\Pi}$ elements and their respective weighted values $\nu(\omega)$ multiplied by $\tilde{\Pi}$ elements, as introduced in eqs.(\ref{CPert}, \ref{Pi_transform1}). All the curves are plotted as a function of the two-photon detuning $\delta=\Delta_2-\Delta_1$.  In all curves the dashed vertical lines show the EIT linewidth of $2.5$MHz, experimentally obtained from the DC signal.

\onecolumngrid

\vspace{.2cm}

\twocolumngrid

Let us first discuss the resonant case  i.e. $\Delta_1=0$, shown in Fig.~\ref{fig:dcHFCorr_signal}b. In this case, we observe a typical EIT profile in the DC signal, which allows the measurement of a linewidth $\Delta\omega_{EIT}$ from the full width at half maximum of the peak.
As for the correlation $C(\omega)$, we notice that the two fields are anti-correlated for $|\delta| > 5$MHz and also for  $|\delta| < 2$MHz, with a narrow peak at exact two-photon resonance. As it was discussed in \cite{Felinto13,HMFlorez13}, this structure is insensitive to power broadening and can be associated to the coherence between the ground states. 
The $g^{(2)}(0)$ function from the heuristic model in eq.(\ref{g20}) is also plotted in solid thin line. Although it presents a similar shape, the result is mostly correlated, with anticorrelation only for  $|\delta| \sim 1$MHz. This is a consequence of the combination of $\text{Im}p_1\text{Im}p_1 $ and $\text{Re}p_1\text{Re}p_1 $ given in eq.~\ref{g20} \cite{Felinto13}. 

In order to understand the observed anticorrelation, we should refer to the terms contributing to eq.~\ref{CPert}. Looking at the values of $\tilde\Pi$ in the EIT region $|\delta| \lesssim  \Delta\omega_{EIT}/2\pi$, we can observe the reduction of the absorptive contribution ($\tilde\Pi_{Im}$) and a competition between the dispersive ($\tilde\Pi_{Re}$) and cross terms ($\tilde\Pi_{IR}$ and $\tilde\Pi_{RI}$). With the weighting factors $\nu(\omega)$, the dominance of dispersion is clear, resulting in the central structure observed with the EIT resonance of the DC signal.
For $|\delta| > 5$MHz, the fields are again anticorrelated since the higher value of $\nu_{IR}$ will compensate the small value of $\tilde{\Pi}_{IR}$ and their product overcomes the other terms in eq.(\ref{CPert}). That means, for $|\delta| \gtrsim \Gamma/2\pi$, the absorption in the fixed beam ($ \text{Im}p_2$) and reemission in the second beam  ($\text{Re}p_1$) dominates in $C$ represented by  $\tilde{\Pi}_{IR}=2\epsilon^2 \text{Re}p_1\text{Im}p_2+\cdots $  i.e. Stokes transitions lead to anticorrelation. 



Similar reasoning can be applied to non-resonant cases (Figs.~\ref{fig:dcHFCorr_signal}a, c), as we can see in the spectroscopic results for different one-photon detunings ($\Delta_1=-6(1)$MHz  and  $5(1)$MHz respectively). 
We can start by observing an interesting feature in the change from correlation to anticorrelation at exact resonance of field 2 ($\Delta_2=0$MHz). The understanding is relatively simple. Fluctuations in the frequency (or equivalently their phase diffusion) are common to both beams, since they are issued from the same laser. 
These fluctuations are converted into intensity fluctuations by a variation in the absorption rate of the fields, as the frequencies approach or recede from resonance.
Therefore when the detuning for both beams have the same sign, intensity fluctuations will be correlated. On the other hand, if detunings have opposite sign, the same correlated fluctuation in phase will lead to anticorrelated response of the medium,  
resulting in anticorrelated fluctuations in intensity. This behavior is closely followed by the dispersive component $\tilde\Pi_{Re}$, that when is weighed by the $\nu_{Re}$ term becomes the leading contributor for most of the spectra.  But contribution of $\tilde\Pi_{RI}$ is also important, specially in the anticorrelated part. Only the contributions of $\tilde\Pi_{Im}$ and $\tilde\Pi_{Re}$, present in the plot of $g^{(2)}(0)$, cannot justify the observed anticorrelation.

Another curious situation shows up in EIT regime, where contribution of dispersion is also reduced: a strong reduction in the correlation occurs in a range of $\pm 3$MHz around the EIT condition $\delta=0$. This structure, carved on the maximized correlation we have just discussed, cannot be completely described only by $g^{(2)}(0)$, since instead of a pair of dips involving a narrow peak we have also another pair of dips close to the analysis frequency (2MHz), composing the complete structure of correlation around EIT condition.
This case is pretty different from the one photon resonant case ($\Delta_1\simeq0$), seen in Fig.~\ref{fig:dcHFCorr_signal}b. Absorptive and dispersive components ($\tilde\Pi_{Im}$ and $\tilde\Pi_{Re}$), after the compensation by $\nu$, are not the leading terms, and the contributions from $\tilde\Pi_{IR}$ and $\tilde\Pi_{RI}$ dominate. Anticorrelation is far from saturation. 

Unfortunately, the structures predicted in theory are not so clearly resolved in the experimental results. Therefore we need a better resolution and a more careful analysis, as we perform next.

\subsection*{Resolving the sideband resonances in the  correlation spectra}\label{sec:Sidebands}

\begin{figure}[htb!]
\centering
\includegraphics[width=86mm]{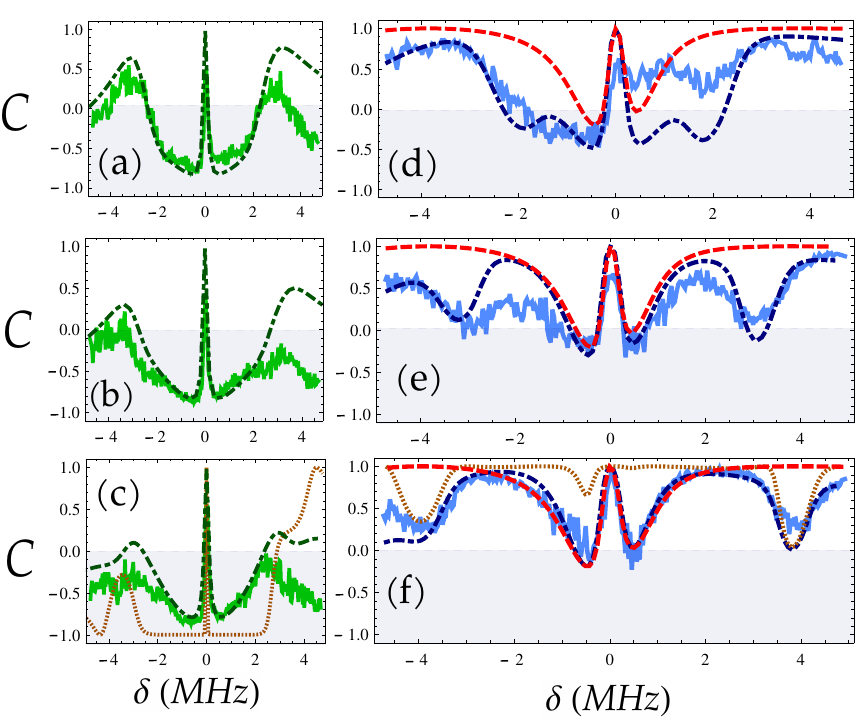}
\vspace{-0.7cm} 
\caption{(Color online) Correlation spectroscopy for different analysis frequencies ($ \omega/2\pi$) with $\Delta_1=0$MHz (first column) and $\Delta_1=6$MHz (second column). (a, d)  $\omega/2\pi=2$MHz, (b, e) $\omega/2\pi=3$MHz and (c, f) $\omega/2\pi=4$MHz. Experimental correlation $C(\omega)$ is in solid line, theoretical result for $C(\omega)$ is traced with dotted-dashed line and $g^{(2)}(0)$ is traced with dashed line. In (c, f) the first order term of the perturbative model is plotted in dotted line.}
\label{fig:Sidebands}
\end{figure}

Looking for a better insight into the role of the sidebands of the fields, we show the correlation spectra for different analysis frequencies $\omega$, at the resonant case $\Delta_1=0$ in Figs.~\ref{fig:Sidebands}(a-c), and also for detuned case $\Delta_1=6$MHz in Figs.~\ref{fig:Sidebands}(d-f). In the first row (Figs.~\ref{fig:Sidebands}(a,d)),  we  have shown the spectra for $\omega/2\pi=2$MHz discussed above.  In the second (Figs.~\ref{fig:Sidebands}(b,e)) and third row (Figs.~\ref{fig:Sidebands}(c,f)) the spectra for $\omega/2\pi=3$MHz and $4$MHz are shown respectively. 

\vspace{-0.0cm}  

The perturbative result of $C(\omega)$ is plotted (dotted-dashed line in Fig.~\ref{fig:Sidebands}) together with the experimental data (solid thick line).   The lowest order term of correlation in $\epsilon^2$  is also plotted in dotted line for $\omega/2\pi=4$MHz.  In  the resonant case $\Delta_1=0$ (see Figs.~\ref{fig:Sidebands}(a-c),  there is no spectral independence for the sidebands in the correlation $C(\omega)$ as discussed in ref.\cite{HMFlorez15}. The intrinsic linewidth of the correlation spectrum (147(9)kHz according to the central peak width) is consistent with the decoherence rate used in the theoretical calculations (150kHz). Although the sideband resonance at $\delta\sim-\omega/2\pi$ seems to be resolved for the first order term (dotted line, Fig.~\ref{fig:Sidebands}c) in the perturbative model, the contribution from higher order terms suppresses the resolution of sideband resonance. This is due to the fact that near resonance (under the natural linewidth of $\sim 6$MHz), the phase-to-amplitude noise conversion process is dominated by the stimulated and spontaneous emission. Thus the excess of phase noise is strongly converted into amplitude noise, populating the sidebands with a smoothly varying distribution. 

Proceeding now with the non-resonant case ($\Delta_1=6$MHz) presented in Figs.~\ref{fig:Sidebands}(d-f),  we notice that the sidebands resonances at $\delta=\pm\omega/2\pi$ are spectrally resolved once that $\omega>\Delta \omega_{EIT}$. For $|\delta|< \Delta \omega_{EIT}/2\pi$, the intense fields (carriers)  determine the correlation profile (Figs.~\ref{fig:Sidebands}(e,f)).  Thus the $g^{(2)}(0)$ overlaps with the correlation $C(\omega)$ near the central structure, where carriers are close to the two-photon resonance. The linewidth of the correlation spectra for $C(\omega)$ and $g^{2}(0)$ are the same (200kHz). 
As for the structure in the range $|\delta| > \Delta \omega_{EIT}/2\pi$, the reduction of the correlation 
at $\delta = \pm \omega/2\pi$ becomes more and more evident for increasing analysis frequencies. The plot of lowest order term in Fig.~\ref{fig:Sidebands}(f) makes evident this effect, absent for $g^{(2)}(0)$.
Since the atomic response to the sideband frequency  only depends on the $\nu(\omega)$ coefficients, the correlation at first order (dotted line) does not seem affected by perturbative corrections in  the sidebands' resonances, differently from the resonant case shown in Fig.~\ref{fig:Sidebands}(c).

 \begin{figure}[htb!]
\centering
\includegraphics[width=8.0cm]{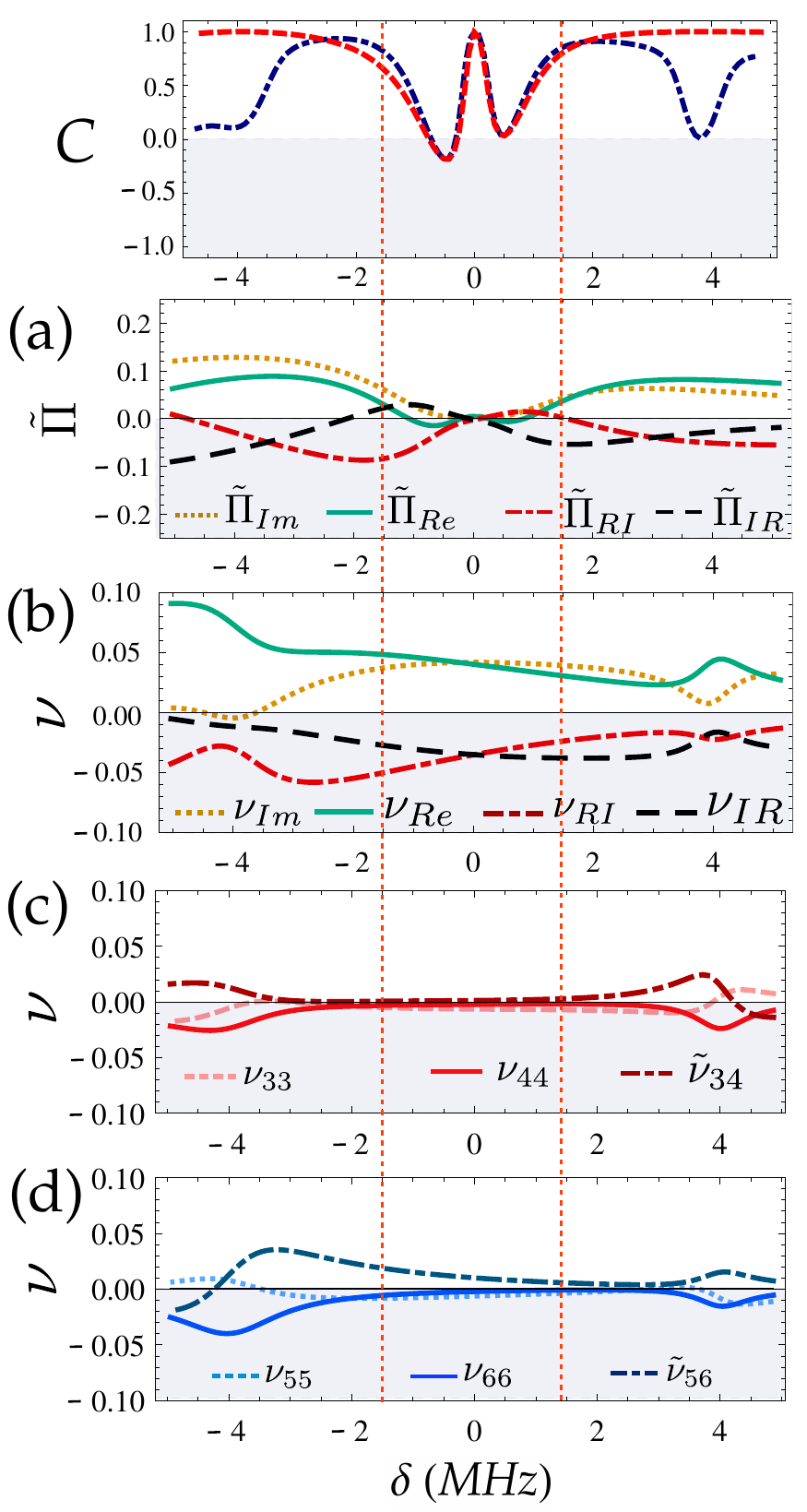}
\vspace{-.4cm}
\caption{(Color online) Mapping between the intensity correlation $C(\omega)$ from Fig.\ref{fig:Sidebands}.(f) and: (a) the $\tilde{\Pi}$ elements and (b) the main $\nu(\omega)$ coefficients. In (c) and (d), the extra $\nu(\omega)$ coefficients that determine $C_1$ defined in eq.(\ref{nu_I2}) are plotted.  This analysis is done for $\Delta_1=5$MHz and $\omega=4$MHz. The two vertical dotted lines indicate the EIT linewidth $\Delta\omega_{EIT}$.}
\label{fig:Coeffabv_wa4MHz}
\end{figure}

In what follows, we will detail the analysis for the case of $\omega/2\pi=4$MHz. 
Figures \ref{fig:Coeffabv_wa4MHz}(a, b) show the main  $\tilde{\Pi}$ elements and $\nu(\omega)$ coefficients, respectively, together with the correlation spectra of Fig. \ref{fig:Sidebands}(f) (redrawn in top row of Fig. \ref{fig:Coeffabv_wa4MHz}). Figures \ref{fig:Coeffabv_wa4MHz}(c, d) show the $\nu(\omega)$ coefficients that define the extra term $C_1$ in eq.(\ref{CPert}) as shown in eq.(\ref{nu_I2}). Unlike the resonant case in Fig. \ref{fig:dcHFCorr_signal}(b) and  Fig. \ref{fig:Sidebands}(c), in  the non-resonant case $\nu_{Im}\sim \nu_{Re}$ for  $|\delta|<3$MHz (see Fig. \ref{fig:Coeffabv_wa4MHz}(a)). Therefore,  the absorptive and dispersive response contribute equally to the intensity correlation, while the term $\tilde\Pi_{cross}$ (eq.~\ref{Pi_transform1}) nearly cancels due to the opposing behavior of $\tilde\Pi_{RI}$ and $\tilde\Pi_{IR}$. This coincides with the heuristic model where the two terms have the same contribution for the $g^{(2)}(0)$ function in eq.(\ref{g20}). The coeficients $\nu_{ij}(\omega)$ for the extra term $C_1$  have almost no contribution to the correlation for $|\delta|<3$MHz, as it is shown in Figs. \ref{fig:Coeffabv_wa4MHz}(c, d).
Hence, the two functions describe the same intensity correlation for  $|\delta|<3$MHz  in the non-resonant case for higher analysis frequency $\omega>\Delta\omega_{EIT}$. 

On the other hand,  the coefficients $\nu(\omega)$ present resonances exactly at the analysis frequency $\delta\sim\pm\omega/2\pi=\pm4$MHz. The main coefficients  $\nu_{Im}(\omega)$ and $\nu_{Re}(\omega)$ present opposite behaviors near the analysis frequency, (Fig. \ref{fig:Coeffabv_wa4MHz}a). The  coefficient $\nu_{Re}$ in solid line shows an increment of the dispersive response  while  $\nu_{Im}(\omega)$ decreases. However, for the sidebands resonance, the contribution of the extra terms in Figs. \ref{fig:Coeffabv_wa4MHz}(c, d) for $C_1$ has also to be considered, leading to $C(\omega)\sim 0$. Curiously, these sideband structures are broad, with nearly half of the width of the EIT peak observed in the DC signal. As we will see, they are also sensitive to power broadening in this case.



\subsection*{Power broadening of the resolved sideband structures}\label{sec:Intensity}

\begin{figure}[htb!]
\centering
\includegraphics[width=88mm]{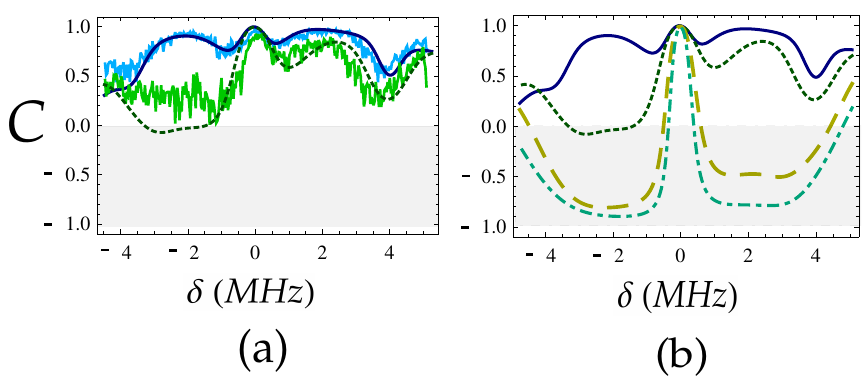}
\vspace{-0.8cm}  
\caption{(Color online) Correlation spectra for different power of the beams in the non-resonant case $\Delta_1=5$MHz, at  4MHz analysis frequency. (a) Experimental correlation spectra for 110$\mu$W  (solid line) and 190$\mu$W (dotted line) with their theoretical counterparts. (b) Theoretical results of the correlation spectra (traced with solid lines) for the powers used in (a), and for higher beam powers, 3$\times$110$\mu$W (dashed line) and 4$\times$110$\mu$W (dotted-dashed lines).}
\label{fig:NoSidebandsPower}
\end{figure}

We further investigated the correlation spectra for different power of the beams at the non-resonant case.
Figure \ref{fig:NoSidebandsPower}a shows the correlation spectra for two different powers of the beams. For beam power of 110$\mu$W, i. e. $I/I_{sat}=0.23$(solid line),  the sideband resonances are well resolved. However, on increasing the power to 190$\mu$W  (dotted line) which corresponds to $I/I_{sat}=0.40$, the resonance at $\delta=-\omega/2\pi$ near the atomic resonance is sensitive to power broadening. 

\vspace{-0.0cm}  

Further increase in the power leads to a significative push of the atoms during the run of the acquisition.  
Nevertheless, we can still consider what are the consequences relying on our theorethical model. Figure \ref{fig:NoSidebandsPower}b shows the theoretical correlation results for the two cases of power of Fig. \ref{fig:NoSidebandsPower}a, and also the calculations for higher beam powers, 3$\times$110$\mu$W (dashed line) and 4$\times$110$\mu$W (dotted-dashed lines) which correspond to $I/I_{sat}=0.67$ and $I/I_{sat}=0.90$, respectively. With increasing power, the central EIT peak broadens and engulfs the sideband structure, now hidden in the anticorrelation flat signal.

\section{Discussion} \label{Discussion}

A linearized approach for the EM field described by eq.~(\ref{Efield}) can give useful hints for the mechanisms involved in the spectroscopy studied here. 
 First, we may regard the phase diffusion as a phase modulation process in the limit of small diffusion coefficient $\gamma$.
 When we measure the spectral components of the photocurrent  $I(t)$ (generated by an intense, nearly monochromatic field) by their Fourier transform $I(\omega)$, we are in fact investigating the beatnotes of the sidebands with the most intense field. For a specific analysis frequency $\omega$ the field of interest will involve three components: the carrier at optical frequency $\omega_i$ and two sidebands shifted by $\pm\omega$. From eq.~(\ref{Efield}) we have the relevant part in calculation of $I(t)$ as
\begin{equation}
\mathbf{E}'_{i}(t)={\cal E}'(t)_{i} \exp (i\omega_{i} t) \mathbf{e}_{i},
\end{equation}
with the complex amplitude given by
\begin{equation}
{\cal E}'(t)_{i}={\cal E}_{i}  +{\cal E}^{u}_{i} \exp (i\omega t)+{\cal E}^{l}_{i} \exp (-i\omega t).
\label{phaseEq}
\end{equation}
in a linearized approach for phase diffusion. Pure amplitude modulation will occur if upper and lower sidebands are symmetric and conjugated (${\cal E}^{u}_{i}=[{\cal E}^{l}_{i}]^*$), and pure phase modulation will take place if they are rotated by $\pi/2$ with respect to the carrier amplitude ${\cal E}_{i} $ (${\cal E}^{u}_{i}=-[{\cal E}^{l}_{i}]^*$).  Phase and amplitude fluctuations can be converted to each other upon asymmetric shifts of the amplitudes of the sidebands (${\cal E}^{u}_{i}$, ${\cal E}^{l}_{i}$) or by a phase shift of the carrier amplitude ${\cal E}_{i}$ \cite{VillarAJP}. In our system, the atoms induce such asymmetric shifts in amplitude and phase.
Photodetection is sensitive to amplitude modulation, therefore we observe in our experiment the conversion of random phase modulation (phase noise) into amplitude (intensity) modulation. 
With that in mind, let us analyse the feature we have observed.

\vspace{-0.0cm}  
\begin{figure}[htb!]
\centering
\includegraphics[width=88mm]{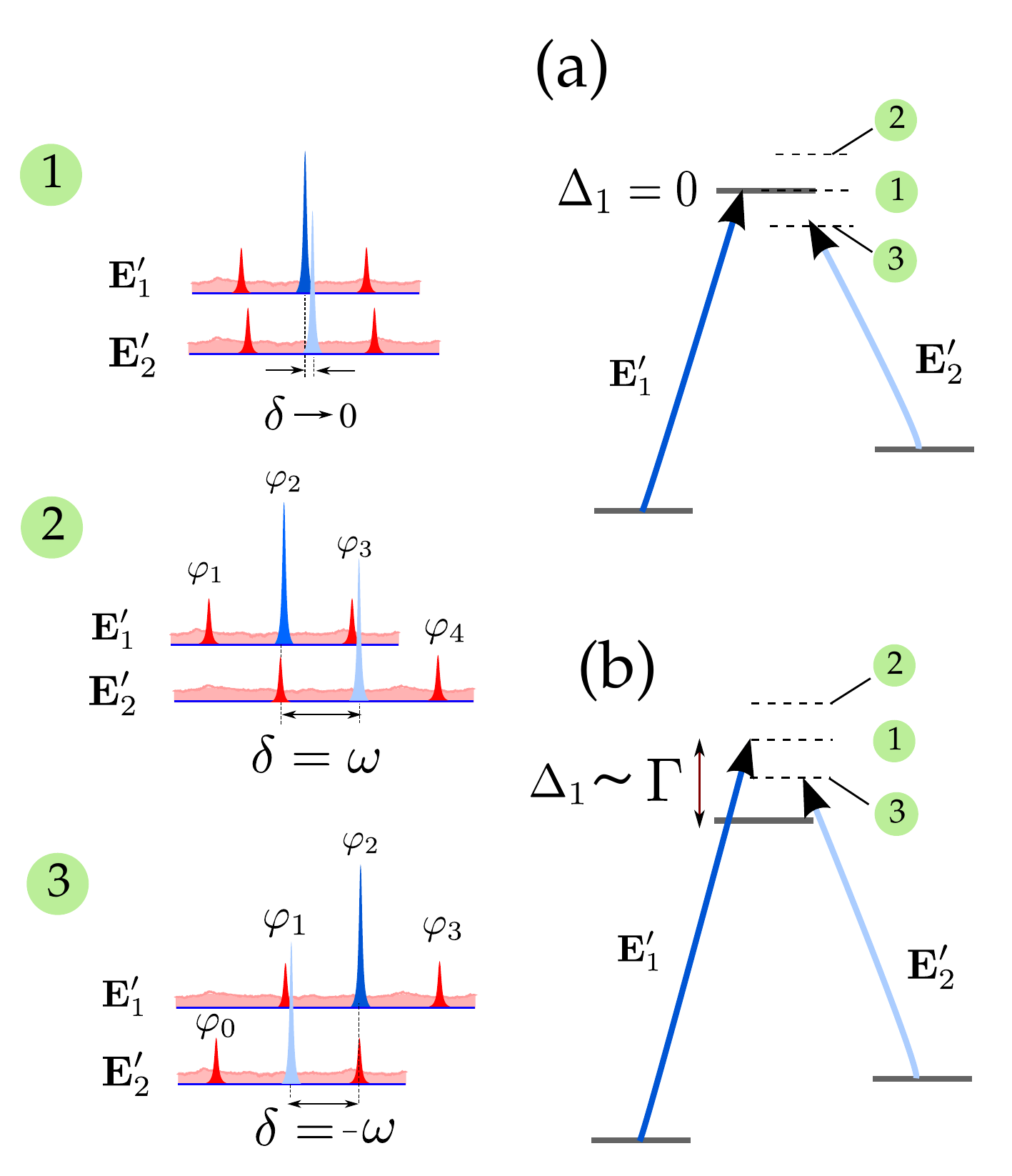}
\vspace{-0.8cm}  
\caption{(Color online) Description for the correlation spectroscopy. Three main situations of interest are: (process 1) the EIT condition  when $\delta\sim 0$, (process 2) when the carrier of the field $\mathbf{E}_2$ is resonant with the upper sideband of the field $\mathbf{E}_1$ ($\delta\sim\omega$) and (process 3) when carrier is resonant with the lower sideband  ($\delta\sim -\omega$).   These processes can be analyzed for two different cases of one-photon detuning:  (a) Resonant case $\Delta_1 = 0$, (b) Non-resonant case $\Delta_1 \sim \Gamma$.}
\label{fig:Cartoon}
\end{figure}

For the case of small single photon detuning ($\Delta_1\simeq 0$), the central structure of $C(\omega)$ that we observe at EIT condition ($|\delta|<\Delta\omega_{EIT}/2\pi$), which is very similar to the $g^{(2)}(0)$ described in \cite{Felinto13}, is a consequence of the phase shifts given to the central carriers while we scan the EIT resonance (see Fig.~\ref{fig:Cartoon}.a). Those shifts convert phase to amplitude noise with opposite signals for fields 1 and 2 (fixed and scanned), leading to a growing anticorrelated response as the intensity is increased, overwhelming the faint intensity correlation that the laser beams may originally present. Only  close to EIT resonance (process 1) those initial intensity correlations are recovered, since the dispersive contribution goes to zero. Therefore we have a narrow central peak in the correlation spectroscopy, either in frequency or time domain. In this case, sidebands are nearly unaffected in the process, or symmetrically affected: upper (lower) sidebands of both beams have similar dephasing and attenuation (similar transformations for ${\cal E}^{u}_{1}$ and ${\cal E}^{u}_{2}$), leading to correlated contributions. The sidebands resonances in the processes 2 and 3 are not resolved due to the stimulated and spontaneous emission, since the fields couple the atoms resonantly with detunings no bigger than the natural linewidth $|\delta|\sim\omega/2\pi<\Gamma/2\pi$ (see figures \ref{fig:Sidebands}(a-c)).

The resonances of the sidebands will be evident in the case when single photon detuning is of the order of the atomic linewidth ($|\Delta_1|\simeq\Gamma/2\pi$, see an example in Fig.~\ref{fig:Cartoon}.b). Once again, close to EIT condition (process 1), sidebands of both beams will suffer similar amplitude and phase changes. Strong correlations are then expected, leading to a maximum of the normalized correlation at exact EIT condition ($|\delta|\simeq 0$) for the same reasons presented in the previous case.

The novelty here is the fact that when detuning matches analysis frequency ($|\delta|=\omega/2\pi$), correlation is reduced. Consider a simple model of phase fluctuations, given in eq.~(\ref{phaseEq}), where ${\cal E}^{u}_{i}=-[{\cal E}^{l}_{i}]^*=\alpha$.
Phase transformation for the fixed beam will be given by:
\begin{eqnarray}
{\cal E}'(t)_{1}&=&\alpha e^{ i\omega t} +{\cal E}_{1}  - \alpha^*  e^{- i\omega t} \rightarrow\\
{\cal E}'(t)_{1}&=&\alpha  e^{ i(\omega t+\varphi_3)} +{\cal E}_{1}  e^{ i\varphi_2}  - \alpha^*_1  e^{ -i(\omega t -\varphi_1)}= \nonumber\\
&=&\left[\alpha  e^{ i(\omega t+\varphi_3-\varphi_2)} +{\cal E}_{1}  - \alpha^*  e^{ -i(\omega t +\varphi_2-\varphi_1)} \right] e^{ i\varphi_2} \nonumber
\end{eqnarray}
Therefore, different phase shifts will convert phase to amplitude fluctuations.

Applying the same transformation for the second field when its carrier is resonant with the upper band of the fixed field (process 2)
\begin{eqnarray}
{\cal E}'(t)_{2}&=&\alpha e^{ i\omega t} +{\cal E}_{2}  - \alpha^*  e^{- i\omega t}\rightarrow \\
{\cal E}'(t)_{2}&=&\alpha  e^{ i(\omega t+\varphi_4)} +{\cal E}_{2}  e^{ i\varphi_3}  - \alpha^*  e^{ -i(\omega t -\varphi_2)}= \nonumber\\
&=&\left[\alpha  e^{ i(\omega t+\varphi_4-\varphi_3)} +{\cal E}_{2}  - \alpha^*  e^{ -i(\omega t +\varphi_3-\varphi_2)} \right] e^{ i\varphi_3} \nonumber.
\end{eqnarray}
Notice that when we discard the overall phase (unobserved upon intensity detection), the effective shift of the upper band of field 1 is the conjugate of the phase shift of the lower sideband of field 2. Fluctuations will rotate in opposite directions in a Fresnel diagram, and resulting intensity correlations will be anticorrelated.

Now if the carrier of the second field is resonant with the lower band of the fixed field (process 3), transformation will be given by
\begin{eqnarray}
{\cal E}'(t)_{2}&=&\alpha e^{ i\omega t} +{\cal E}_{2}  - \alpha^*  e^{- i\omega t}\rightarrow \\
{\cal E}'(t)_{2}&=&\alpha  e^{ i(\omega t+\varphi_2)} +{\cal E}_{2}  e^{ i\varphi_1}  - \alpha^*  e^{ -i(\omega t -\varphi_0)}= \nonumber\\
&=&\left[\alpha  e^{ i(\omega t+\varphi_2-\varphi_1)} +{\cal E}_{2}  - \alpha^*  e^{ -i(\omega t +\varphi_1-\varphi_0)} \right] e^{ i\varphi_1} \nonumber.
\end{eqnarray}
The phase shift of the upper sideband of field 2 will be the conjugate of the phase shift of the upper sideband of field 1. Once again, anticorrelation will contribute to the result of $C(\omega)$.

This conversion depends on the phase gained by each mode. Under power broadening of the EIT resonance \cite{Felinto13,HMFlorez13}, anticorrelation will be affected by the reduction on the diference of phase shifts, as is the case in Fig.~\ref{fig:NoSidebandsPower}.

\section{Conclusions}\label{Conclusions}

Correlation spectroscopy for two light fields interacting with cold atomic medium at different one-photon detunings has been explored.
When the one-photon detuning is of the order of the natural linewidth and beam intensities are much smaller than the saturation limit, the correlation spectroscopy presents resonances at the sidebands frequency. If the analysis frequency is larger than the typical EIT linewidth, the intensity correlation spectra in the EIT range can be approximately described by the $g^{(2)}(0)$ function \cite{Felinto13}.
Therefore, the intrinsic  linewidth measured with the  $g^{(2)}(0)$ function in time domain corresponds exactly to the one observed with the correlation spectrum $C(\omega)$ in frequency domain. 

Sidebands resonances are auxiliar spectroscopic tools for the study of EIT process, and can be resolved for analysis frequencies higher than the EIT linewidth. However, in such a condition, the sidebands resonances are sensitive to power broadening.  It was also shown that for fields close to atomic resonance, the contribution from the sidebands and the carriers are mixed, disabling any spectral independence of sidebands in the correlation spectroscopy. The resolution of the sidebands resonances is only possible with the use of cold atoms. Next step is to measure correlation spectra  using a coherent source to verify if it is possible to measure a narrower intrinsic linewidth for resonant conditions as it is predicted by the first order term in the perturbative model \cite{HMFlorez15}.

This work was supported by grant \# 2010/08448-2, S\~ao Paulo Research Foundation (FAPESP),  CNPq and CAPES (Brazilian agencies), through the programs PROCAD, PRONEX, and INCT-IQ (Instituto Nacional de Ci\^encia e Tecnologia de Informa{\c c}\~ao Qu\^antica). The authors would like to thank Prof. Dr. Vanderlei S. Bagnato and Dr. Kilvia M. F. Magalh\~aes, from IFSC-USP, for gently providing the Rb for our MOT.
                                                    
                                                                                                                                                                                                                                                                                                                                                                                                                                                                                   \appendix

\section{Solution for Noise spectra}\label{apendA}

The intensity correlation result in eq.(\ref{CPert}) is demonstrated in detail in ref.\cite{HMFlorez15}. Shortly, to describe the role of the  atomic response  in the PN-AN conversion, we should begin by writing the density matrix elements  $\mathbf{x}=(\rho_{11},\rho_{22},\rho_{13},\rho_{31},\rho_{23},\rho_{32},\rho_{12},\rho_{21})$. It can be expanded on the $\epsilon$ parameter, that we will eventually associate with the laser linewidth,
\begin{align}
 \mathbf{x}(t) = \mathbf{x}^{(0)}(t) + \epsilon\ \mathbf{x}^{(1)}(t) + \epsilon^2 \mathbf{x}^{(2)}(t) + \cdots\ \label{ExpanPert}
\end{align}
The stationary state of the zero order terms is defined as $\langle \mathbf{x}_{ss}^{(0)}\rangle=(\rho_{11}^{(0)},\rho_{22}^{(0)},p_1,p_1^*,p_2,p_2^*, ,\rho_{12}^{(0)},\rho_{21}^{(0)})$. Its stationary solution is  $\langle \mathbf{x}_{ss}^{(0)}\rangle=\mathbf{M}^{-1}\mathbf{x}_o$ with $\mathbf{x}_o$ as a constant vector and  the $\mathbf{M}$ matrix contains all the parameters of the atom-light interaction.

The covariances between the atomic density matrix elements are also expanded as 

\begin{align}
\Pi =&\epsilon^2\sigma^{(0)} +\sum_{n=2}^\infty \epsilon^{2n} \sigma^{(2n-2)},\label{Pi_expand}
\end{align}
where $\Pi_{ij}= \langle \mathbf{x}_{i}(t),\mathbf{x}_j(t)^\dagger \rangle$ and $\sigma_{ij}= \langle \mathbf{x}_{i}^{(n)}(t),\mathbf{x}_j^{(n)}(t)^\dagger \rangle$. The stationary solution for the zero order term is defined as
\begin{align}
\sigma_{ss}^{(0)}=(\mathbf{M})^{-1} \mathbf{x}_o \mathbf{x}_o^\dagger (\mathbf{M}^\dagger)^{-1}. \label{sigma_0}
\end{align}

It is useful to transform the atomic vector as $\tilde{\mathbf{x}}(t)=\mathbf{U}\ \mathbf{x}(t)$ where $\mathbf{U}$ converts $x_i(t)$ for $i=\{3-6\}$ into their real and imaginary part.
Therefore the covariances $\Pi$ and $\sigma^{n}$ will transform also according to $\mathbf{U}$  as $\tilde{\Pi}$ and $\tilde{\sigma}^{(n)}$.

The noise spectra  $S_{11}$ and $S_{22}$ in eq.(\ref{CPert}) are given by
 \begin{align}
S_{11}(\omega) =&\alpha_{33}(\omega)\left[2\epsilon^2  \text{Im}^2p_1 + \sum_{n=2}^\infty \epsilon^{2n}\tilde{\sigma}_{33}^{(2n-2)}\right]\nonumber \\&+\alpha_{44}(\omega)\left[2\epsilon^2  \text{Re}^2p_1 + \sum_{n=2}^\infty \epsilon^{2n}\tilde{\sigma}_{44}^{(2n-2)}\right]\nonumber \\
&+\tilde{\alpha}_{34}(\omega)\left[-2\epsilon^2 \text{Im}p_1\text{Re}p_1 +\sum_{n=2}^\infty \epsilon^{2n}\tilde{\sigma}_{34}^{(2n-2)} \right]\nonumber \\&+\alpha_{C}(\omega)  +\alpha_{I2}(\omega), \label{s11_pert_final}
\end{align}
and 
\begin{align}
S_{22}(\omega) =& \beta_{55}(\omega)\left[2\epsilon^2 \text{Im}^2p_2 + \sum_{n=2}^\infty \epsilon^{2n}\tilde{\sigma}_{55}^{(2n-2)}\right]\nonumber\\&+\beta_{66}(\omega)\left[2\epsilon^2  \text{Re}^2p_2 + \sum_{n=2}^\infty \epsilon^{2n}\tilde{\sigma}_{66}^{(2n-2)}\right]\nonumber \\
&+\tilde{\beta}_{56}(\omega)\left[-2\epsilon^2 \text{Im}p_2\text{Re}p_2 +\sum_{n=2}^\infty \epsilon^{2n}\tilde{\sigma}_{56}^{(2n-2)} \right]\nonumber \\&+\tilde{\beta}_{C}(\omega) + \beta_{I1}(\omega), \label{s22_pert_final}
\end{align}
where $\tilde{O}_{ij}(\omega)=O_{ij}(\omega)+ O_{ji}(\omega)$ with $O=\alpha,\ \beta$ and $\nu$. The products $ \text{Im} p_i\text{Im}p_j$, $ \text{Re} p_i\text{Re}p_j$ and $ \text{Im}p_i\text{Re}p_j$ are associated to the elements $\sigma^{(0)}_{i+2,j+2}$ with $i,j=1,2$. 
The additional terms are defined as 
\begin{align}
\alpha_{I_2} &= \alpha_{55}(\omega)\tilde{\Pi}_{55} +\alpha_{66}(\omega)\tilde{\Pi}_{66} +\tilde{\alpha}_{56}(\omega)\tilde{\Pi}_{56} \label{alpha_I2}\\
\alpha_{C} &= \tilde{\alpha}_{35}(\omega)\tilde{\Pi}_{35} +\tilde{\alpha}_{46}(\omega)\tilde{\Pi}_{46} \nonumber\\&+ \tilde{\alpha}_{36}(\omega)\tilde{\Pi}_{36}+\tilde{\alpha}_{45}(\omega)\tilde{\Pi}_{45}\\
\beta{I_1}&= \beta_{33}(\omega)\tilde{\Pi}_{33} +\beta_{44}(\omega)\tilde{\Pi}_{44} +\tilde{\beta}_{34}(\omega)\tilde{\Pi}_{34}\\
\beta_{C}&= \tilde{\beta}_{35}(\omega)\tilde{\Pi}_{35} +\tilde{\beta}_{46}(\omega)\tilde{\Pi}_{46}\nonumber\\&+ \tilde{\beta}_{36}(\omega)\tilde{\Pi}_{36}+\tilde{\beta}_{45}(\omega)\tilde{\Pi}_{45}
\end{align}

The cross product that defines the intensity correlation in eq.(\ref{CPert}) is given by
\begin{align}
S_{12}(\omega) =& \nu_{Im}(\omega)\left[2\epsilon^2  \text{Im} p_1\text{Im}p_2 + \sum_{n=2}^\infty \epsilon^{2n}\tilde{\sigma}_{35}^{(2n-2)}\right]\nonumber \\&+\nu_{Re}(\omega)\left[2\epsilon^2  \text{Re}p_1\text{Re}p_2 + \sum_{n=2}^\infty \epsilon^{2n}\tilde{\sigma}_{46}^{(2n-2)}\right]\nonumber \\
&+\nu_{RI}(\omega)\left[-2\epsilon^2 \text{Im}p_1\text{Re}p_2 +\sum_{n=2}^\infty \epsilon^{2n}\tilde{\sigma}_{36}^{(2n-2)} \right]\nonumber \\&+\nu_{IR}(\omega)\left[-2\epsilon^2 \text{Im}p_2\text{Re}p_1 +\sum_{n=2}^\infty \epsilon^{2n}\tilde{\sigma}_{45}^{(2n-2)} \right]\nonumber\\
&+C_1,\label{s12_pert_final}
\end{align}
where $C_1$ is given by

\begin{align}
C_1 = & \nu_{33}(\omega)\tilde{\Pi}_{33} +\nu_{44}(\omega)\tilde{\Pi}_{44} +\tilde{\nu}_{34}(\omega)\tilde{\Pi}_{34}\nonumber \\
&+ \nu_{55}(\omega)\tilde{\Pi}_{55} +\nu_{66}(\omega)\tilde{\Pi}_{66} +\tilde{\nu}_{56}(\omega\tilde{\Pi}_{56}\label{nu_I2}
\end{align}

\end{document}